\documentclass[a4paper]{spie}  
\usepackage[dvips]{graphicx}

\def\CN2{\mbox{$C_N^2 \ $}}

\def\CT2{\mbox{$C_T^2 \ $}}

\def\sigmal2{\mbox{$\sigma ^{2}_{I} \ $}}

\title{Optical turbulence: site selection above the internal antarctic plateau with a mesoscale model} 

\author{Elena Masciadri\supit{a}, Franck Lascaux\supit{a}, Susanna Hagelin\supit{a,b}, 
\skiplinehalf
\supit{a}INAF - Osservatorio Astrofisico di Arcetri, L.go E. Fermi 5, 50125  Florence, Italy
\skiplinehalf
\supit{b}Departement of Earth Sciences, Uppsala University, Uppsala, Sweden\\
}

\authorinfo{Correspondence to masciadri@arcetri.astro.it; lascaux@arcetri.astro.it}

\begin{document} 
  \maketitle 

\begin{abstract}
Atmospherical mesoscale models can offer unique potentialities to characterize and discriminate potential astronomical sites.
Our team has recently completely validated the Meso-Nh model above Dome C (Lascaux et al. 2009, 2010). Using all the measurements of \CN2 profiles (15 nights) performed so far at Dome C during the winter time (Trinquet et al. 2008) we proved that the model can reconstruct, on rich statistical samples, reliable values of all the three most important parameters characterizing the turbulence features of an antarctic site: the surface layer thickness, the seeing in the free atmosphere and in the surface layer. Using the same Meso-Nh model configuration validated above Dome C, an extended study is now on-going for other sites above the antarctic plateau, more precisely South Pole and Dome A.   
In this contribution we present the most important results obtained in the model validation process and the results obtained in the comparison between different astronomical sites above the internal plateau. The Meso-Nh model confirms its ability in discriminating between different optical turbulence behaviors, and there is evidence that the three sites have different characteristics regarding the seeing and the surface layer thickness.
We highlight that this study provides the first homogeneous estimate, done with comparable statistics, of the optical turbulence developed in the whole 20-22 km above the ground at Dome C, South Pole and Dome A. 
\end{abstract}



\section{INTRODUCTION}
\label{sec:intro}  

The Internal Antarctic Plateau represents a potential interesting location for astronomical applications. Since almost a decade astronomers showed more and more interest towards this region of the Earth thanks to its peculiar atmospheric conditions. The extreme cold temperature, the dry atmosphere, the fact that  the plateau is at more than 2500 m from the sea level, the turbulence seems to develop mainly in a thin surface layer of the order of 30-40 m on the top of summits and the seeing above this surface layer assumes values comparable to those obtained at mid-latitude sites get this region of the earth very appealing for astronomers. South Pole has been the first site equipped with an Observatory in the Internal Antarctic Plateau in which measurements of the optical turbulence have been done (Marks et al. 1996, Marks et al. 1999). Fifteen balloons have been launched in the winter period and it has been observed that the seeing above a surface layer of $\sim$ 220 m was very good (0.37 arcsec). Measurements of the optical turbulence at Dome C are more recent. After the first observations done in 2004 with a MASS (Lawrence et al. 2004), appeared a series of studies done with different instrumentation that supported the assessment of the integrated seeing (Aristidi et al. 2005, Aristidi et al. 2009) and the vertical distribution of the optical turbulence (Trinquet et al. 2008). 

\begin{table*}
\caption{Results obtained in Lascaux et al. 2010 that proved the Meso-Nh model reliability above Dome C in reconstructing the optical turbulence spatial distribution. Three parameters are estimated: the mean surface turbulent 
  layer (h$_{sl}$), the seeing in the free atmosphere ($\varepsilon_{FA}$) obtained integrating the \CN2 from h$_{sl}$ up to the end of the atmosphere, the total seeing ($\varepsilon_{TOT}$) obtained integrating the \CN2 from the ground up the top of the atmosphere. Beside each parameter is reported the associated standard deviation ($\sigma$) and the statistical error ($\sigma$/$\sqrt{N}$).}
\begin{center}
\begin{tabular}{cccccccccc}
   \hline
        & h$_{sl}$    & $\sigma$    & $\sigma$/$\sqrt{N}$   &  $\varepsilon_{FA}$   &    $\sigma$    & $\sigma$/$\sqrt{N}$ & $\varepsilon_{TOT}$   &    $\sigma$    & $\sigma$/$\sqrt{N}$\\
        &       (m)      & &                            &    (arcsec)     & &                    & (arcsec)  & & \\
  \hline
 Observations     & 35.3 &  19.9 & 5.1 & 0.30 & 0.70 & 0.20 & 1.60 & 0.70 & 0.20 \\  
 Model  & 44.2 &  24.6 & 6.6 & 0.30 &0.67 &0.17 & 1.70 & 0.77 & 0.21 \\
   \hline
 \end{tabular}
 \end{center}
\label{tab_sum}
\end{table*}

In this paper we treat a different approach to the site assessment. In this context we are interested in investigating the abilities of a mesoscale model (Meso-Nh) in reconstructing correct optical turbulence features above different sites of the Internal Antarctic Plateau and its abilities in discriminating  the optical turbulence properties of different sites. Meso-Nh (Lafore et al., 1998) is a non-hydrostatic mesoscale research model developed jointly by the Centre National des Recherches M\'et\'eorologiques (CNRM) and the Laboratoire d'Aereologie de Toulouse, France. The Astro-Meso-NH package (Masciadri et al. 1999a) has been proved for the first time to be able to reconstruct realistic $\CN2$ profiles above astronomical sites by Masciadri et al. (1999b) and Masciadri et al. (2001) and statistically validated later on (Masciadri \& Jabouille, 2001, Masciadri et al. 2004, Masciadri et al. 2006). More recently, it has been statistically validated above Dome C by Lascaux et al. (2009, 2010). The most important results obtained in these last papers are summarized in Table \ref{tab_sum}.

Briefly, the observations at Dome C, for a set of 15 winter nights (all the available nights for which is known the optical turbulence vertical distribution), gave a mean surface layer thickness $h_{sl,obs}$ = 35.3 $\pm$ 5.1 m.
The simulated surface layer thickness obtained with the Meso-NH model ($h_{sl,mnh}$ = 44.2 $\pm$ 6.6 m) is well correlated to measurements. The statistical error is of the order of 5-6 m but the standard deviation ($\sigma$) is of the order of 20-25 m. This indicates that the statistic fluctuation of this parameter is intrinsically quite important.
The median simulated free-atmosphere seeing ($\varepsilon_{mnh,FA}$ = 0.30 $\pm$ 0.17 arcsec) as well as the the median total seeing ($\varepsilon_{mnh,TOT}$ = 1.70 $\pm$ 0.21 arcsec) are well correlated to observations, respectively $\varepsilon_{obs,FA}$ = 0.3 $\pm$ 0.2 arcsec and $\varepsilon_{obs,TOT}$ = 1.6 $\pm$ 0.2 arcesc.

In the context of this paper we consider that the Meso-Nh model is calibrated as shown in Lascaux et al. (2010) i.e. it 
produces optical turbulence features in agreement with observations. We therefore apply the Meso-Nh 
model with the same configuration to other two sites of the plateau: South Pole and Dome A (Table \ref{tab0}). 

\begin{table*}
\caption{Geographic coordinates of Dome A, Dome C and South Pole. The altitude is in meter.}
\begin{center}
\begin{tabular}{lrrcc}
\hline
 SITE          & LATITUDE             & LONGITUDE           & MESO-NH      & MEASURED    \\
               &                      &                     & ALTITUDE (m) & ALTITUDE (m) \\
  \hline
 Dome A$^*$    & 80$^{\circ}$22'00"S  &077$^{\circ}$21'11"E & 4089 & 4093 \\  
 Dome C$^{**}$ & 75$^{\circ}$06'04"S  &123$^{\circ}$20'48"E & 3230 & 3233 \\
 South Pole    & 90$^{\circ}$00'00"S  &000$^{\circ}$00'00"E & 2746 & 2835 \\ 
   \hline
\multicolumn{4}{l}{$^*$ {\scriptsize GPS measurement by Dr. X. Cui (private communication).}}\\
\multicolumn{4}{l}{$^{**}$ {\scriptsize GPS measurement by Prof. J. Storey (private communication).}}
\end{tabular}
\end{center}
 \label{tab0}
\end{table*}

Why these sites ? Dome A is an almost uncontaminated site of the plateau. It is the highest summit of the plateau and, for this reason, it is expected to be among the best astronomical sites for astronomical 
applications. The high altitude reduces the whole atmospheric path for light coming from space and above the summit the katabatic wind speed is reduced to minima values. Dome A has been proved to have the strongest thermal stability (Hagelin et al. 2008) in proximity of the ground due to the coldest temperature. Dome A is a chinese base. In the last few years the chinese astronomers gave a great impulse to the 
site characterization showing a great interest for building astronomical facilities in this site. Optical turbulence measurements during the winter time are not yet available.
South Pole is interesting in our study because measurements of optical turbulence are available and, at the same time, the site is not located on a summit but above a gently slope. From the preliminary measurements done in the past we expect a surface turbulent layer that is thicker than the surface layer developed above the other two sites (Dome C and Dome A) due to the ground slope and the consequent katabatic winds in proximity of the surface. The three sites form therefore a perfect sample for a benchmark test on the model behavior and the model abilities.

\begin{table}
\caption{Meso-NH model configuration. In the second column the  horizontal resolution $\Delta$X, in the third column the number of grid points and in the fourth column the horizontal surface covered by the model domain.}
\begin{center}
\begin{tabular}{cccc}
\hline
Domain & $\Delta$X & Grid Points & Surface \\
 & (km) & & (km$\times$km) \\
\hline
Domain 1       & 25& 120$\times$120& 3000$\times$3000\\
Domain 2 & 5 &  80$\times$80&   400$\times$400 \\
Domain 3  & 1 &  80$\times$80& 80$\times$80  \\
\hline
\end{tabular}
\end{center}
\label{tab2}
\end{table}

In Section \ref{num} the numerical set-up of the model is presented. In Section \ref{opt} results of the complete analysis of the three major parameters that characterize the optical turbulence features: surface layer thickness, seeing in the free atmosphere i.e. calculated above the surface layer and total seeing are reported. Two different criteria to define the surface layer are used with consequent double treatment. Finally, in Section \ref{concl} the results of this study are summarized.

\section{NUMERICAL SET-UP}
\label{num}
A detailed description of the Meso-NH mesoscale model \cite{laf} used for this study is presented in \cite{lf09,lf10}.
We briefly recall here the main characteristics of the numerical configuration:
\begin{itemize}
\item The interactive grid-nesting technique \cite{st} is used, with three imbricated domains of increased horizontal mesh-sizes 
($\Delta$X=25 km, 5 km and 1 km, Table \ref{tab2}). Such a method is used to permit us to achieve the best resolution on a small surface but keeping the volumetric domain in which the simulation is done in thermodynamic equilibrium with the atmospherical circulation that evolves at large spatial scale on larger domains.
\item The vertical grid is the same for all the domains reported in Table \ref{tab2}. The first vertical grid point is at 2 m above ground level (a.g.l.). A logarithmic stretched grid 
up to 3500 m above ground level (a.g.l.) (with 12 points in the first hundred of meters) is employed. Above 3500 m a.g.l., the vertical resolution is constant ($\Delta$H $\sim$ 600 m). The maximum altitude achieved is around 22 km a.g.l..
\item All simulations are initialized and forced every 6 hours at synoptic times (00:00, 06:00, 12:00, 18:00) UTC by analyses from the European 
Center for Medium-range Weather Forecasts (ECMWF)\footnote{ECMWF: $http://www.ecmwf.int/$}. Note that the time at which the simulation starts (UTC) differs for Dome A, Dome C and South Pole. This is done so to be able to compare optical turbulence profiles simulated in the same temporal interval with respect to the local time (LT). 
For each night, a mean vertical profile of \CN2 is computed between the time interval (20:00 - 00:00) LT as does in Lascaux et al. (2009, 2010).
In Table \ref{tab1} are reported, for each site, the time at which the simulation starts and the duration $\Delta$T of the simulation with respect to the local time.
\item An optimized version of the externalized surface scheme ISBA (Interaction Soil Biosphere Atmosphere) for antarctic 
conditions is employed \cite{lem,lem10}. Such a scheme has been used in Lascaux et al. 2010 and it contributed to provide a realistic reconstruction of the optical turbulence near the surface (optical turbulence strength and turbulence layer thickness). It is indeed obvious that the most critical part of an atmospherical model for this kind of simulations is the scheme that controls the air/ground turbulent fluxes budget. Our ability in well reconstructing the surface temperature T$_{s}$ is related to the ability in reconstructing the sensible heat flux H that is responsible of the buoyancy-driven turbulence in the surface layer. 
\item The Astro-Meso-NH package (Masciadri et al. 1999) implemented in the most recent version of Meso-Nh has been used to calculate the optical turbulence and derived astroclimatic parameters. 
\end{itemize}

\begin{table*}
\caption{Simulation starting time and time interval chosen for \CN2 computations for the 3 different sets of simulations 
           (Dome A, Dome C and South Pole).}
\begin{center}
\begin{tabular}{cccc}
   \hline
                     & Dome A               & Dome C               & South Pole            \\
   \hline
Starting time        & 06:00 UTC / 11:00 LT & 00:00 UTC / 08:00 LT & 12:00 UTC / 12:00 LT  \\
   \hline
Time interval        &                      &                      &                       \\
for \CN2 computations & 15:00 - 19:00 UTC    & 12:00 - 16:00 UTC    & 20:00 - 00:00 UTC     \\
(20:00 - 00:00 LT)   &                      &                      &                       \\
   \hline
\end{tabular}
\end{center}
\label{tab1}
\end{table*}

\begin{figure*}
\begin{center}
\includegraphics[width=\textwidth]{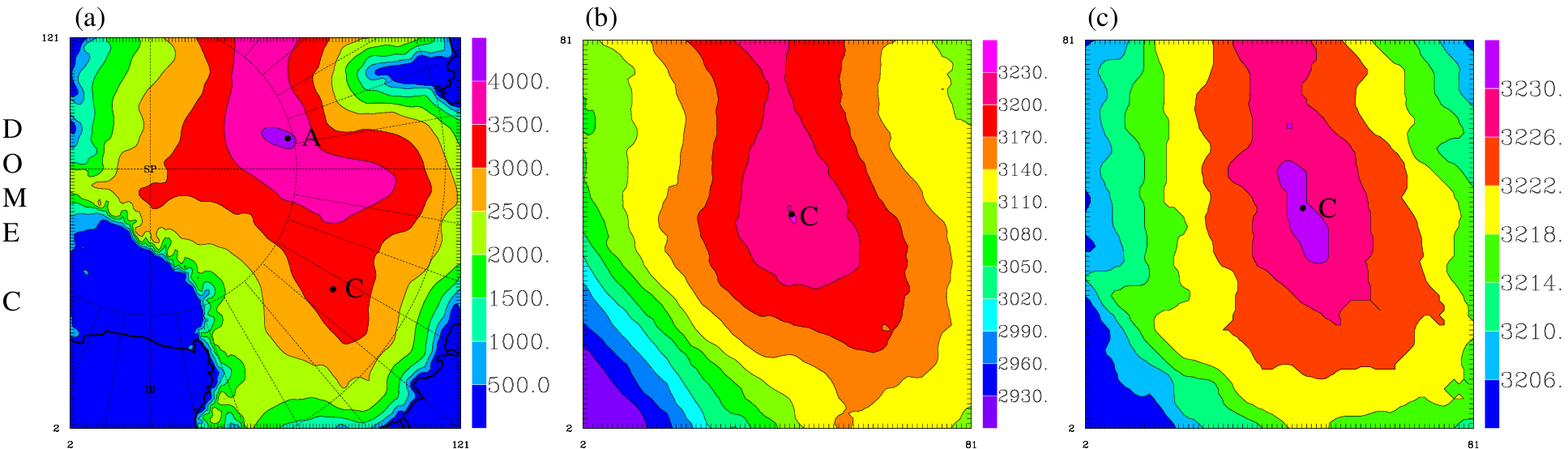}
\includegraphics[width=\textwidth]{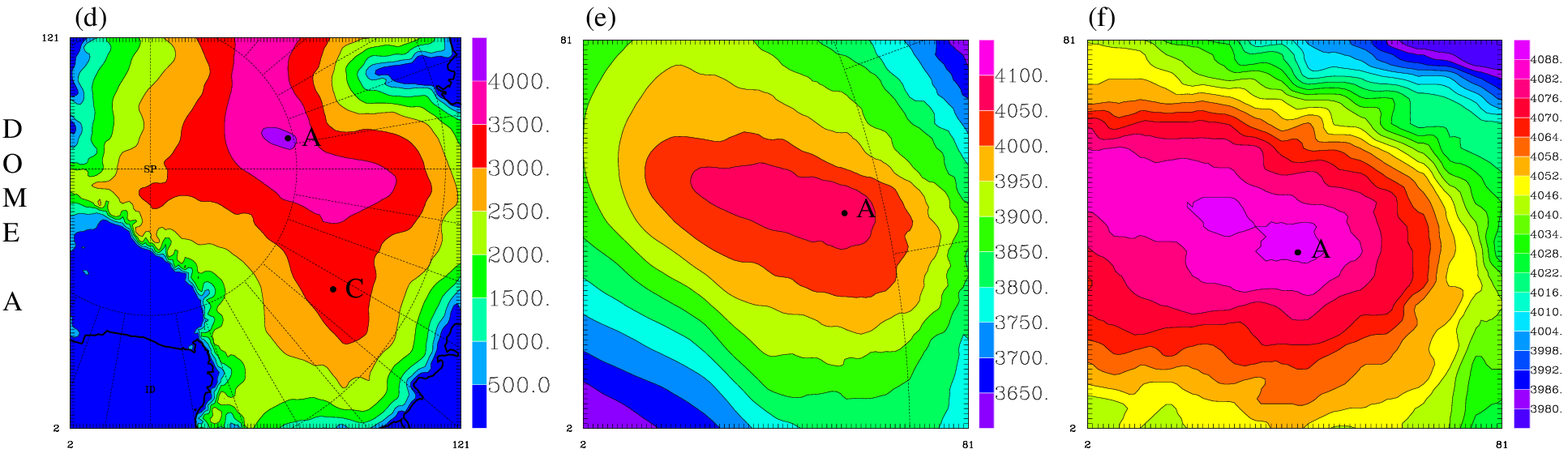}
\includegraphics[width=\textwidth]{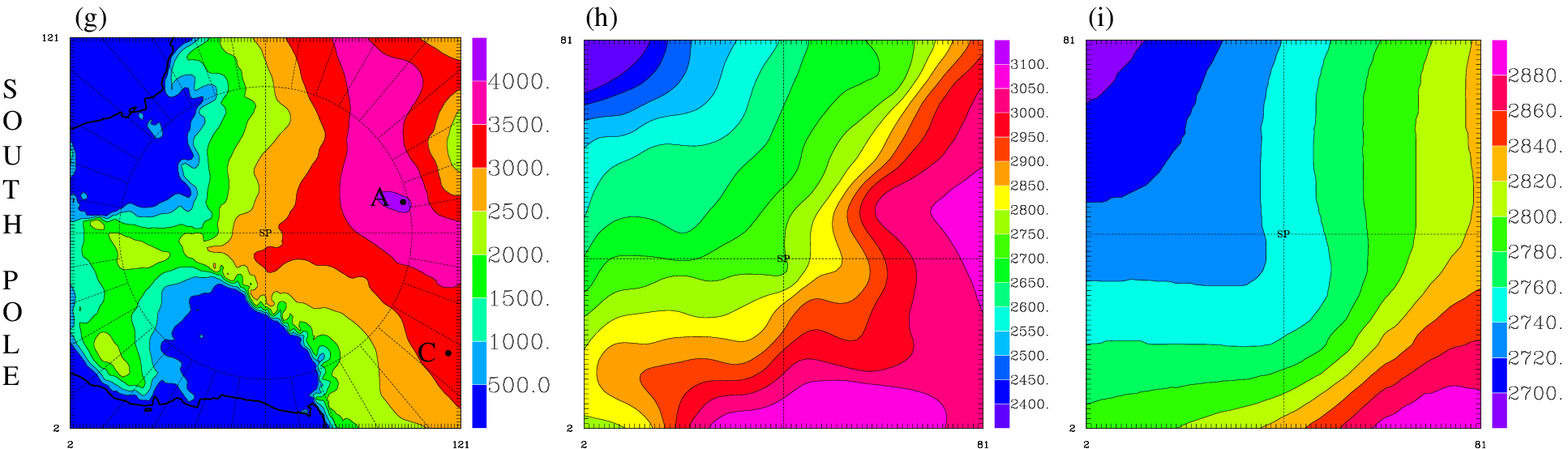}
\end{center}
\caption{Orography of three different regions of the internal Antarctic Plateau as seen by the Meson-Nh model (polar
 stereographic projection, grid-nesting configuration).
(a), (b) and (c) show the three imbricated domains for the Dome C simulations, with horizontal resolution of 25 km, 5 km and 1 km, respectively.
(d), (e) and (f) show the three imbricated domains for the Dome A simulations, with horizontal resolution of 25 km, 5 km and 1 km, respectively.
(g), (h) and (i) show the three imbricated domains for the South Pole simulations, with horizontal resolution of 25 km, 5 km and 1 km, 
respectively.
The dot labeled 'C' indicates the Concordia Station. The dot labeled 'A' indicates Dome A. SP stands for South Pole. 
The altitude is expressed in meter (m).}
{\label{fig1}}
\end{figure*}

As shown in \cite{lf10}, the best choice for the description of the orography is the RAMP (Radarsat Antarctic Mapping Project) 
Digital Elevation Model (DEM) presented in \cite{liu}, instead of the GTOPO30 DEM from the U.S. Geological Survey used in \cite{lf09}.
For this study, therefore, the RAMP Digital Elevation Model has been used.  The orography of each area of interest in this study (Dome C, Dome A, South Pole) is displayed on Fig.~\ref{fig1}. All the grid-nested 
domains, from low horizontal resolution (larger mesh-size) to high horizontal resolution (smaller mesh-size) are displayed.
As can be seen in Fig.~\ref{fig1} (c,f,i) the orography around Dome C and Dome A is more detailed than the orography in proximity of the South Pole. This is due to the fact that the procedure to obtain a DEM integrates data from many different sources (satellite radar altimetry, airborne surveys, 
GPS surveys, station-based radar sounding...). However the resolution of some areas (typically those that can hardly receive information from the satellites) remain poorer than others. The region included in the inner circular polar region (and therefore South Pole) fits with this condition and this is the reason why the orography is somehow less detailed than the rest of the Internal Antarctic Plateau. 
Nevertheless, this is a region with no peaks or mountains and with just a regular and gently slope. 
We can therefore reasonably expect that the poorer accuracy in the orography has little or minor influence on the results of 
the numerical simulations done with a mesoscale model such as Meso-NH.\\
The same set of 15 winter nights used by \cite{lf09,lf10} to validate the model above Dome C is investigated in this study for the three antarctic sites 
Dome C, Dome A and South Pole. 

\section{OPTICAL TURBULENCE ABOVE DOME C, DOME A, SOUTH POLE}
\label{opt}
In this section we investigate and compare the values obtained above the three sites (Dome C, Dome A and South Pole) of three parameters that characterize the optical turbulence features above the antarctic plateau:
\begin{itemize}
\item surface layer thickness;
\item free atmosphere seeing from the the surface layer thickness (h$_{sl}$) up to the top of the atmosphere;
\item total seeing from the ground up to the top of the atmosphere. We note that this corresponds to $\sim$10 km above ground level because the baloons explode at this altitude due to the high pressure and the strong wind speed.
\end{itemize}

\subsection{Optical Turbulence surface layer thickness}
To compute the surface layer thickness for each night, the same method employed in \cite{tr} and \cite{lf10} is first used. 
The thickness $h_{sl}$ is defined as the vertical slab containing 90 per cent of the optical turbulence developed inside 
the first kilometer above the ground:
\begin{equation}
 \label{eq:bl1}
 \frac{ \int_{8m}^{h_{sl}} C_N^2(h)dh }{ \int_{8m}^{1km} C_N^2(h)dh } < 0.90
\end{equation}
where $C_N^2$ is the refractive index structure parameter. We remind here that the selection of this criterium (that we call criterium A) is motivated by the fact that we intend 
to compare our calculations with measurements done by Trinquet et al. (2008). This criterium has been selected by Trinquet et al. (2008) because the typical optical turbulence features above the internal antarctic plateau is characterized by a major bump at the surface and a consistent decreasing of the optical turbulence strength in the first tens of meters. The selection of the percentage is obviously absolutely arbitrary and it assumes a value only when the same criterium is applied in other sites. 

\begin{table}
 \caption{Mean surface layer thicknesses $h_{sl}$ computed for the 3 sites, for 15 different winter nights using the criterion in
          Eq. \ref{eq:bl1}. Units in meter (m). The mean values are also reported  with the associated 
          statistical error $\sigma$/$\sqrt{N}$.}
 \begin{center}
 \begin{tabular}{crrr}
 \hline
 Date                & Dome A & Dome C & South Pole \\
 \hline
 04/07/05            &  65.0  &  30.4  & 117.6      \\ 
 07/07/05            & 529.4  &  35.4  & 262.9      \\
 11/07/05            &  28.6  &  80.0  & 131.9      \\
 18/07/05            &  27.7  &  49.7  & 224.0      \\
 21/07/05            &  17.6  &  66.7  & 136.3      \\
 25/07/05            &  15.7  &  27.4  & 298.6      \\
 01/08/05            &  25.5  &  22.6  & 185.2      \\
 08/08/05            &  53.1  &  34.2  & 104.4      \\
 12/08/05            &  19.4  &  16.7  &  59.0      \\
 29/08/05            &  17.4  &  91.4  & 251.3      \\ 
 02/09/05            &  16.4  &  70.9  & 164.9      \\
 05/09/05            & 125.2  & 338.4  & 128.0      \\  
 07/09/05            &  59.8  &  52.5  & 103.6      \\
 16/09/05            &  38.8  &  19.4  & 158.7      \\
 21/09/05            &  20.1  &  21.0  & 148.0      \\
 \hline
 Mean                &  37.9* &  44.2* & 165.0      \\
 \hline
 $\sigma$            &  30.2* &  24.6* &  67.2      \\
 \hline
 $\sigma$/$\sqrt{N}$ &   8.1* &   6.6* &  17.4      \\
 \hline
\multicolumn{4}{l}{*These values are computed without taking into account the} \\
\multicolumn{4}{l}{night of the 05/09/05 for Dome C and the night of 07/07/05} \\
\multicolumn{4}{l}{for Dome A (see text for further explanations).} \\
 \end{tabular}
 \end{center}
 \label{tab3}
\end{table}

Table \ref{tab3} reports the computed values of the surface layer thickness for each night at the three sites, 
as well as the mean, the standard deviation ($\sigma$) and the statistical error ($\sigma/\sqrt N$) for the 15 nights.
For each night, the surface layer thickness is computed from a computed \CN2 profile averaged between 20 LT and 00 LT (see Table \ref{tab2} 
for hours in UT) as done in Lascaux et al. 2010. The calculated mean surface layer thickness above South Pole (h$_{sl}$$=$165 m $\pm$ 14.3 m) is more than three time larger than that developed above Dome C (h$_{sl}$$=$43.5 m $\pm$ 5.9 m) and Dome A (h$_{sl}$$=$44.2 m $\pm$ 8.2 m). This difference is well correlated with previous observations done above South Pole. More precisely, observations related to 15 balloons launched during the period (20/6/1995 - 18/8/1995) indicated h$_{sl}$$=$220 m \cite{ma3}. Measurements in that paper are done in winter but in a different year and different nights. It is not surprising therefore that the matching between calculations and measurements is not perfect. Unfortunately the precise dates of nights studied in the paper from Marks et al. (1999) are not known. It is therefore not possible to provide a more careful estimate. 
It is however, remarkable the fact that the h$_{sl}$ above South Pole is substantially larger than the h$_{sl}$ above Dome C and Dome A. 
Also we note that the typical thickness calculated above South Pole with a statistical sample of three months by \cite{seg} was h$_{sl}$$=$102 m.  
The authors used however a different definition of turbulent layer thickness. More precisely, they defined h$_{sl}$ as the elevation (starting from the lowest model level) at which the turbulent kinetic energy contains 1 per cent of the turbulent kinetic energy of the lowest model layer. A comparison of this result with our calculations and with measurements is therefore meaningless. The same conclusion is valid for the estimates of h$_{sl}$ given at Dome C as already explained in Lascaux et al. (2009, 2010). In conclusion, looking at Table \ref{tab3}, individuals values for each nights show a $h_{sl,SP}$ almost always higher than 100 m, with a maximum 
close to 300 m (2005 July 25), whereas $h_{sl,DC}$ and $h_{sl,DA}$ are always below 100 m.
Dome C and Dome A have a comparable surface layer thickness. 
For this sample of 15 nights, $h_{sl,DA}$ is 6.3 m smaller than $h_{sl,DC}$. We note also that the number of nights for which $h_{sl}$ is 
very small (inferior at 30 m) is more important at Dome A (nine instead of six at Dome C). This difference is however not really statistically reliable considering the number of the nights in the sample. For a more detailed discrimination between the h$_{sl}$ value at Dome C and Dome A we need a larger statistic. This analysis is planned for a forthcoming paper.

Looking at the results obtained night by night we can note some specific features observe in specific cases. Two nights (September 5 at Dome C ($h_{sl}$ = 338.4 m) and July 7 at Dome A ($h_{sl}$ = 529.4 m)) present similar characteristics: the surface layer thickness h$_{sl}$ is well larger than the observed one. In these two cases, however, as already explained in \cite{lf09,lf10} for the Dome C site, the large value of h$_{sl}$ does not mean that a thicker and more developed turbulence is present near the ground but it simply means that, in the first kilometer from the ground, 90 per cent of the turbulence develops in the (0, h$_{sl}$) range. In both cases the model reconstructs the total seeing on the whole 20 km much weaker than what has been observed and more uniformly distributed and, consequently, the criterium (Eq.\ref{eq:bl1}) provides us a much larger value of h$_{sl}$. 
In both cases,  when we look at the vertical distribution of the \CN2, we observe that the turbulence is concentrated well below 20 m in a very thin surface layer with a very weak total seeing (see next section and table \ref{tab4}).
The case of 5 September at Dome A, is however a case in which the model reconstructed a surface turbulent layer thicker than what has been observed.

In order to compare our calculations and results with those obtained by \cite{seg} we applied also a different criterium (criterium B) based on the analysis of the  vertical profile of turbulent kinetic energy (TKE) instead of the vertical profile of \CN2. The TKE is certainly an ingredient from which the optical turbulence depends on and it represents the dynamic turbulent energy. However, it is known (Masciadri \& Jabouille 2001) that the $\CN2$ depends also on the gradient of the potential temperature and moreover, it is absolutely arbitrary the selection of the value of percentage of the turbulent kinetic energy (1$\%$, 10$\%$, other...) used as a threshold. This method is not therefore useful to quantify the absolute value of h$_{sl}$ to be compared to measurements provided by Trinquet et al. (2008) and Marks et al. (1999). It can be, eventually, useful for relative comparisons between different sites or to compare our calculations with calculations provided by Swain \& Gallee (2006).

Using this method (Table \ref{tab_TKE}), the surface layer height is determined as the elevation at which the TKE is X\% of the lowest elevation value. We calculated the h$_{sl}$ for X = 1 (Table \ref{tab_TKE}) and X = 10 (Table \ref{tab_TKEbis}). X=1 is the case treated by \cite{seg}. For each simulation, we first compute the average of the TKE profile for the night between 20 LT and 00 LT. While the average of the $\CN2$ profile is calculated with a 2 minutes rate sample, the average of the TKE is calculated with 5 profiles, available at each hour (20, 21, 22, 23, 24, 00) LT.  This gives us an averaged vertical profile of TKE characteristic of the considered night. 

The computation of the surface layer thickness is then performed using this averaged TKE profile.
It has been observed that, when the night presents only low dynamic turbulence (with a very low averaged TKE at the 
lowest elevation level), it is very hard to retrieve a surface layer height using this criterium. This means that the turbulence is so weak that we are at the limit of necessary turbulent kinetic energy to resolve the turbulence itself.
For these nights (indicated with an asterisk in Table \ref{tab_TKE}) it could happen that we calculated the average on a number of estimates minor than 5 (as for all the other cases). The results are reported in Table \ref{tab_TKE}.

\begin{table}
 \caption{Mean surface layer thicknesses $h_{sl}$ computed for the 3 sites, for the same set of nights shown in Table \ref{tab3}, but 
          computed with a different criterion. The surface layer height is determined as the elevation at which the averaged TKE 
          between 20 LT and 00 LT for each nigh is 1\% of the averaged lowest elevation value. Units in meter (m). The mean values are also 
          reported  with the associated statistical error $\sigma$/$\sqrt{N}$.}
 \begin{center}
 \begin{tabular}{crrr}
 \hline
 Date                & Dome A & Dome C & South Pole \\
 \hline
 04/07/05            &    78* &    32  &   112      \\
 07/07/05            &    6* &    32  &   112*     \\
 11/07/05            &    40  &    76  &   174      \\
 18/07/05            &    32  &    48  &   242      \\
 21/07/05            &    22  &    56  &   144      \\
 25/07/05            &    22  &    12* &   148      \\
 01/08/05            &    32  &    22  &   186      \\
 08/08/05            &    56  &    32  &   112      \\
 12/08/05            &    22  &    60  &    58      \\
 29/08/05            &    22  &    82  &   250      \\
 02/09/05            &    20  &    60  &   188      \\
 05/09/05            &   136  &    30* &   146      \\
 07/09/05            &    72  &    74  &   250      \\
 16/09/05            &    56  &    22  &   192      \\
 21/09/05            &    26  &    22  &   170      \\
 \hline
 Mean                &    42.8&    44&   165.6    \\
 \hline
 $\sigma$            &    33.0&    22.6  &    55.5    \\
 \hline
 $\sigma$/$\sqrt{N}$ &     8.5&     5.8&    14.3    \\
 \hline
\multicolumn{4}{l}{*These values are computed using a number } \\
\multicolumn{4}{l}{of profiles minor than five.} \\
 \end{tabular}
 \end{center}
 \label{tab_TKE}
\end{table}

Table \ref{tab_TKE} shows that results obtained with the criterium of the TKE are similar to those obtained with the criterium described in Eq.\ref{eq:bl1}. Table \ref{tab_TKEbis} provides, however smaller values of h$_{sl}$ above all the three sites. We treat the case (X = 10) to show that, tuning the value of the percentage, it is possible to find different values of h$_{sl}$. This means that estimates are useful only if they are compared to measurements using the same criteria.  

\begin{table}
 \caption{Mean surface layer thicknesses $h_{sl}$ computed for the 3 sites, for the same set of nights shown in Table \ref{tab3}, but 
          computed with a different criterion. The surface layer height is determined as the elevation at which the averaged TKE 
          between 20 LT and 00 LT for each nigh is 10\% of the averaged lowest elevation value. Units in meter (m). The mean values are also 
          reported  with the associated statistical error $\sigma$/$\sqrt{N}$.}
 \begin{center}
 \begin{tabular}{crrr}
 \hline
 Date                & Dome A & Dome C & South Pole \\
 \hline
 04/07/05            &    58 &    22  &   56      \\
 07/07/05            &    2 &    24  &   62     \\
 11/07/05            &    28  &    52  &   64      \\
 18/07/05            &    22  &    32  &   186      \\
 21/07/05            &    14  &    40  &   110     \\
 25/07/05            &    12  &    6 &   102     \\
 01/08/05            &    22  &    16  &   126      \\
 08/08/05            &    40  &    24  &   102      \\
 12/08/05            &    16  &    8  &    40      \\
 29/08/05            &    14  &    68  &   192      \\
 02/09/05            &    14  &    46  &   112      \\
 05/09/05            &   106  &    30 &   88      \\
 07/09/05            &    50  &    52  &   68      \\
 16/09/05            &    40  &    12  &   150      \\
 21/09/05            &    18  &    14  &   116     \\
 \hline
 Mean                &    30.4 &    27&   104.9    \\
 \hline
 $\sigma$            &    26.1&    15.6  &    45.2    \\
 \hline
 $\sigma$/$\sqrt{N}$ &     6.7&     4&    11.7    \\
 \hline
 \end{tabular}
 \end{center}
 \label{tab_TKEbis}
\end{table}

\begin{table}
\caption{Total seeing $\varepsilon_{TOT}$$=$$\varepsilon_{[8m,h_{top}]}$ and seeing in the free atmosphere
$\varepsilon_{FA}$$=$$\varepsilon_{[h_{sl},h_{top}]}$ calculated for the 15 nights and averaged in the temporal range
20-00 LT .
See the text for the definition of h$_{sl}$ and h$_{top}$.
In the second column are reported the observed values, in the third and fourth columns the simulated values
obtained with high and low horizontal resolution respectively. Units in arcsec.}
\begin{center}
\begin{tabular}{cccc}
\hline
& DOME A      & DOME C               & SOUTH POLE         \\
\hline
  Date     & $\varepsilon_{FA}$/$\varepsilon_{TOT}$      & $\varepsilon_{FA}$/$\varepsilon_{TOT}$      & $\varepsilon_{FA}$/
$\varepsilon_{TOT}$        \\
           & {\tiny($h_{sl}$=37.9m)} & {\tiny($h_{sl}$=44.2m)} & {\tiny($h_{sl}$=165m)} \\
 \hline
  04/07/05 &      2.55 / 3.37      &     0.22 / 2.28      &    0.40 / 1.67         \\
  07/07/05 &      0.20 / 0.24      &     0.28 / 1.91      &    0.31 / 0.70         \\
  11/07/05 &      0.23 / 2.78      &     1.61 / 1.81      &    0.47 / 1.96         \\
  18/07/05 &      0.21 / 2.73      &     0.80 / 1.94      &    1.46 / 2.28         \\
  21/07/05 &      0.21 / 1.95      &     0.86 / 1.27      &    0.31 / 1.71         \\
  25/07/05 &      0.22 / 1.55      &     0.25 / 0.85      &    0.32 / 0.76         \\
  01/08/05 &      0.22 / 1.78      &     0.22 / 2.27      &    0.52 / 1.78         \\
  08/08/05 &      1.45 / 2.42      &     0.35 / 1.70      &    0.28 / 1.69         \\
  12/08/05 &      0.23 / 2.37      &     0.23 / 0.99      &    0.29 / 1.82         \\
  29/08/05 &      0.23 / 1.83      &     2.29 / 2.47      &    1.55 / 2.11         \\
  02/09/05 &      0.22 / 1.76      &     1.16 / 1.54      &    0.81 / 3.56         \\
  05/09/05 &      3.21 / 3.36      &     0.30 / 0.52      &    0.31 / 2.98         \\
  07/09/05 &      2.43 / 3.49      &     1.69 / 3.73      &    0.31 / 1.41         \\
  16/09/05 &      1.11 / 4.60      &     0.21 / 1.57      &    0.99 / 3.96         \\
  21/09/05 &      0.20 / 2.30      &     0.26 / 1.63      &    0.36 / 2.32         \\
 \hline
  Median   &      0.23 / 2.37      &     0.30 / 1.70      &    0.36 / 1.82         \\
 \hline
  $\sigma$ &      1.08 / 1.03      &     0.67 / 0.77      &    0.43 / 0.90         \\
 \hline
  $\sigma$/$\sqrt{N}$ & 0.28 / 0.27&     0.17 / 0.21      &    0.11 / 0.23         \\
 \hline
\end{tabular}
\end{center}
\label{tab4}
\end{table}

To conclude, both criteria (A and B with X=1) give similar mean $h_{sl}$ values for all the 3 sites for this limited set of nights.
Evaluating the surface layer thickness over a more extended set of nights should be the next step. It would permit us to compute more 
reliable and robust statistical estimates for $h_{sl}$ over the 3 antarctic sites. 
This also means that our estimate (h$_{sl}$=165 m) above South Pole is better correlated to measurements (h$_{sl}$=220 m) that obtained by \cite{seg} (h$_{sl}$=102 m) at the same site. 

\subsection{Seeing in the free atmosphere and seeing in the whole atmosphere}

\begin{figure*}
\begin{center}
\includegraphics[width=\textwidth]{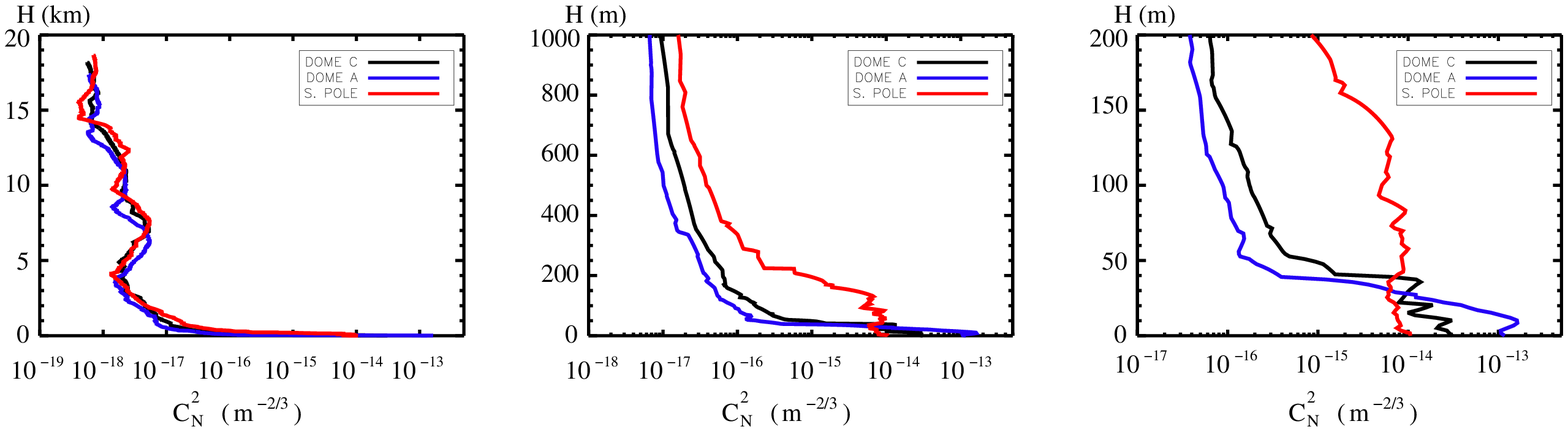}
\end{center}
\caption{Median \CN2 profiles simulated with the Meso-NH mesoscale model at Dome C 
(black), Dome A (blue) and South Pole (red). Left: from the ground up to 20 km. Middle: 
from the ground up to 1 km. Right: from the ground up to 200 m. Units are $m^{-2/3}$.}
\label{fig2}
\end{figure*}

Table \ref{tab4} shows the simulated total seeing ($\varepsilon_{TOT}$) and free-atmosphere seeing ($\varepsilon_{FA}$) for 
each night and each sites (Dome C, Dome A and South Pole). We define the free atmosphere as the portion of the atmosphere extended from the mean $h_{sl}$ reported in Table \ref{tab3} up to 10 km. At this height the balloon explode in general during the winter and we need to limit the integration of the $\CN2$ at such a low value to be able to compare our results with those obtained with measurements.
The median values of the seeing as well as the standard deviation ($\sigma$) and the statistical error ($\sigma/\sqrt N$) are reported.
As expected the total seeing is a little bit stronger at Dome A ($\varepsilon_{TOT,DA}$ = 2.37 $\pm$ 0.27 arcsec) than at Dome C 
($\varepsilon_{TOT,DC}$ = 1.70 $\pm$ 0.21 arcsec) or South Pole ($\varepsilon_{TOT,SP}$ = 1.82 $\pm$ 0.23). 
The total seeing is very well correlated with measurements at Dome C (Lascaux et al. 2010 - $\varepsilon_{TOT,obs}$ = 1.6 arcsec) and at South Pole (Marks et al. 1999 - $\varepsilon_{TOT,obs}$ = 1.86 arsec) getting the estimate at Dome A highly reliable.
The minimum median free-atmosphere seeing at Dome A is $\varepsilon_{FA,DA}$ = 0.23 $\pm$ 0.28 arcsec, at Dome C, $\varepsilon_{FA,DC}$ = 0.30 $\pm$ 0.17 arcsec and at South Pole, $\varepsilon_{FA,SP}$ = 0.36 $\pm$ 0.11 arcsec. The seeing in the free atmosphere is very well correlated with measurements at Dome C (Lascaux et al. 2010 - $\varepsilon_{FA,obs}$ = 0.30 arcsec) and at South Pole (Marks et al. 1999 - $\varepsilon_{FA,obs}$ = 0.37 arsec) getting again very reliable the method (Meso-NH model) as well as the estimates at Dome A.
What is remarkable is that, even if $h_{sl,DA}$ $<$ $h_{sl,DC}$ $<$ $h_{sl,SP}$, Dome A is the site with the lowest free-atmosphere seeing $\varepsilon_{FA}$. This means that at Dome A as well as at Dome C the turbulence is concentrated inside the first tenths of meters 
from the ground. Moreover, the turbulence in the surface layer is stronger at Dome A than at Dome C. This can be explained with the 
stronger thermal stability of Dome A near the ground. Our results match, therefore, with predictions we did in Hagelin et al. (2008) studying 
only features of the meteorological parameters. At South Pole, however, the \CN2 vertical distribution decreases in a less abrupt way.
and it is spread over hundredth of meters from the ground,
instead of tenths of meters like for Dome A or Dome C. As a result the total seeing is also 
weaker than above Dome C and Dome A. Such a behavior is evidenced in Figure \ref{fig2}, which displays the median vertical
\CN2 profiles over the 3 sites.

\section{CONCLUSION}
\label{concl}
In this study the mesoscale model Meso-NH was used to perform forecasts of optical turbulence (evolutions of \CN2 profiles) 
for 15 winter nights at three different antarctic sites: Dome A, Dome C and South Pole. The model has been used with the same configuration 
previously validated at Dome C (Lascaux et al. 2010) and simulations of the same 15 nights have been performed above the three sites. The idea behind our approach is that once validated above Dome C, the model can be used above two other site above the internal antarctic plateau to discriminate optical turbulence features typical of other sites. This should show the potentiality of the numerical tool in the context of the site selection and characterization in astronomy.  South Pole has been chosen because in the past some measurements of the optical turbulence have been done and this can represent a useful constraint for the model itself. For Dome A there are not at present time measurements of the optical turbulence and this study provides therefore the first estimates ever done of the optical turbulence above this site. 
We test this approach above the antarctic plateau because this region is particularly simple from the morphologic point of view and certainly simpler than typical mid-latitude astronomical sites. No major mountain chains are present and the local surface circulations is mainly addressed by the energy budget air/ground transfer, the polar vortex circulation at synoptic scale and the katabatic winds generated by gravity effects on gently slopes due to the cold temperature of the iced surface. 

The main results we obtained are:
\begin{itemize}
\item The Meso-Nh model achieves to reconstruct the three most important parameters used to characterize the optical turbulence: the turbulent surface layer thickness, the seeing in the free atmosphere and in the surface layer for the three selected sites: Dome C, Dome A and South Pole showing results in agreement with expectations. Measurements taken at Dome C and South Pole corresponds to balloons launched during 15 nights, in both cases. The statistic is not very large but reliable for a first significant result. The selected nights correspond to the 15 nights for which measurements of the Dome C are available. 
\item Dome C and Dome A present a very thin surface layer size ($h_{sl,DA}$ = 37.9 $\pm$ 8.1 m and  $h_{sl,DC}$ = 44.2 $\pm$ 6.6 m) while South Pole surface layer is much thicker ($h_{sl,SP}$ = 165 $\pm$ 17.4 m). All the estimates are well correlated with measurements. To better discriminate between Dome A and Dome C surface layer thickness a richer statistic is necessary. An on-going study has started addressing this issue.
\item Dome A is the site with the strongest total seeing (2.37 $\pm$ 0.27 arcsec) with respect to Dome C ($\varepsilon_{TOT,DC}$ = 1.70 $\pm$ 0.21 arcsec) and South Pole ($\varepsilon_{TOT,SP}$ = 1.82 $\pm$ 0.23 arcsec). This is explained by the stronger thermal stability near the ground with respect to other two sites that cause large values of the optical turbulence in the thin surface layer.
\item All the three sites show a very weak seeing in the free atmosphere i.e. above the correspondent mean h$_{sl}$: $\varepsilon_{FA,DA}$ = 0.23 $\pm$ 0.28 arcsec at Dome A,  $\varepsilon_{FA,DC}$ = 0.30 $\pm$ 0.17 arcsec at Dome C and $\varepsilon_{FA,SP}$ = 0.36 $\pm$ 0.11 arcsec at South Pole. Dome A show the weakest seeing in the free atmosphere. 
\item Both, the total seeing and the seeing in the free atmosphere calculated by Meso-Nh, are very well correlated with measurements at Dome C and South Pole getting the predictions done at Dome A highly reliable.
\item Dealing on the criteria used to define the surface layer thickness, we proved that, at least on the sample of 15 nights investigated, the criterium defined by Eq.\ref{eq:bl1} (criterium A) and the criterium using the vertical profile of the turbulent kinetic energy (TKE) taking h$_{sl}$ as the height at which the value of the TKE is less than 1$\%$ of the TKE at the lowest level near the ground (criterium B) provide more or less the same results. 
\item The mean h$_{sl}$ we estimate at Dome C (h$_{sl}$=44.2) is slightly thicker than what found by \cite{seg} (h$_{sl}$=27.7 m) with comparable discrepancy from measurements (h$_{sl}$ = 35.3 $\pm$ 5.1 m). The h$_{sl}$ we estimate at South Pole ($h_{sl,SP}$ = 165 $\pm$ 17.4 m) is thicker than what estimated by \cite{seg} ($h_{sl,SP}$ = 102) but better correlated to measurements ($h_{sl,SP}$ = 220 m) than what found by \cite{seg}. The h$_{sl}$ we estimate at Dome A ($h_{sl,DA}$ = 37.9 $\pm$ 8.1 m) is somehow thicker than what estimated by \cite{seg} ($h_{sl,DA}$ = 18 m). It is however important to note that the standard deviation of h$_{sl}$ is of the order of h$_{sl}$ itself or even larger. The statistic error $\sigma$/$\sqrt(N)$ is of the order of $\sim$ 10 m. We think therefore that at present there are no major differences in our results with respect to \cite{seg} with expection of the fact that we proved that, with our model, the horizontal resolution of 1 km provides better results than a resolution of 100 km (\cite{seg}).
\end{itemize}
All these results deserve now a confirmation provided by an analysis done with a richer statistical sample. Besides, we can state that all major expectations concerning the typical features of the optical turbulence above South Pole, Dome C and Dome A have been confirmed by this study. 
We highlight that this paper provides the first estimate of the optical turbulence extended on the whole 20 km above the Internal Antarctic Plateau.

\acknowledgments     
 
ECMWF products are extracted from the catalogue MARS,
http://www.ecmwf.int, access to these data was authorized by the
Meteorologic Service of the Italian Air Force. This study has been
funded by the Marie Curie Excellence Grant (FOROT) -
MEXT-CT-2005-023878.

\end{document}